\documentclass[fleqn,twoside]{article}
\usepackage{espcrc2}

\newcommand{\beq}{\begin{equation}}
\newcommand{\eeq}{\end{equation}}
\newcommand{\beqs}{\begin{eqnarray}}
\newcommand{\eeqs}{\end{eqnarray}}

\usepackage{graphicx}
\usepackage[figuresright]{rotating}

\newcommand{\AmS}{{\protect\the\textfont2
  A\kern-.1667em\lower.5ex\hbox{M}\kern-.125emS}}

\title{Gravitational duality in General Relativity and Supergravity theories.}

\author{F. Dehouck \address[MCSD]{Service de physique math\'ematique et interactions fondamentales. \\ Universit\'e Libre de Bruxelles, Campus Plaine CP-231, 1050 Bruxelles, Belgium}%
        }

\begin{document}

\begin{abstract}
We quickly review the current status of gravitational duality in General Relativity. We summarize and comment  some recent work on constructing dual (topological) charges and understanding how this  duality acts in supergravity theories.
\vspace{1pc}
\end{abstract}

\maketitle

\section{Introduction}

It is a well known fact that General Relativity's equations, when reduced to three dimensions, possess an hidden $SL(2,R)$ symmetry known as Ehler's symmetry.
Gravitational duality refers to an $SO(2)$ subgroup of it. In this sense, it was first understood as a symmetry of the space of solutions with a Killing vector. The famous example is the Schwarzschild solution, with mass $M$, that is mapped under duality to the Taub-NUT solution, with NUT charge $N$. Surprisingly, the story is not over as it was recently shown in \cite{Henneaux:2004jw} that
the Pauli-Fierz Hamiltonian, describing the four-dimensional linearized theory, also possesses this  symmetry, without any reference to the presence of Killing vectors. However, the duality is still very  hypothetical at the non-linear level in the absence of Killing vectors (See also \cite{Deser:2005sz}).

\section{Gravitational duality for linearized general relativity}

In the linearized theory, gravitational duality interchanges the equation of motion with the cyclic identity. The Bianchi identity stays invariant. This is obvious when the equations are written as (see for example \cite{Argurio:2009xr} and references therein):
\beqs
G^{(a)}_{\mu\nu}=T^{(a)}_{\mu\nu},
\eeqs
where $a=1,2$ describes the electric or magnetic linearized Einstein tensor and we introduced a magnetic stress-energy tensor in the cyclic identity. The Riemann tensors are related by:
\beqs
 R^{(2)}_{\mu\nu\rho\sigma}=\frac{1}{2} \epsilon_{ \mu\nu\alpha\beta} R^{(1) \:\alpha\beta}_{\qquad \rho\sigma} ,
\eeqs
and generic $SO(2)$ transformations of those tensors, together with rotations of the stress-energy tensors, keep the equations of motion invariant.  This symmetry was made manifest at the level of the action in \cite{Henneaux:2004jw} where the existence of a generator of duality and its associated conserved Noether charge was unveiled.

This work was generalized in the presence of a cosmological constant in \cite{Julia:2005ze}. In this case, the duality relates the on-shell electric and magnetic parts of the Weyl tensor. One recovers the results of the asymptotically flat case in the limit $\Lambda \rightarrow 0$.

\section{Existence of dual charges}

As magnetic charges are by definition topological, one can not proceed through the Noether procedure to define them. In \cite{Argurio:2009xr}, we defined ten additionnal dual charges. These were postulated in analogy with the analysis of electromagnetic duality. Dual momenta were recovered  by a Witten-Nester construction in \cite{Argurio:2008zt} \footnote{  See also \cite{Barnich:2008ts} and  \cite{Bossard:2008sw} }.  The twenty charges are ($8\pi G=1$):
\beqs
P_{\mu}&=& \int G^{(1)}_{0\mu} d^3 x^{\mu} ,\;\: L_{\mu\nu}=2\int G^{(1)}_{0[\mu} x_{\nu]} d^3x^{\mu}  ,\nonumber \\
K_{\mu}&=& \int G^{(2)}_{0\mu} d^3 x^{\mu} , \;\: \tilde{L}_{\mu\nu}=2\int G^{(2)}_{0[\mu} x_{\nu]} d^3x^{\mu} ,
\eeqs
and we also found a way to express them as surface integrals.  Looking at the Kerr-NUT solution, we find $P_0=M,  K_0=N, L_{xy}=Ma, \tilde{L}_{xy}=Na$ when appropriate delta singularities are taken into account. It is clear, however, that  a proper treatment of the Lorentz charges in GR should include information about non-linearities.  In the common approach, usual (electric) Poincar\' e charges are recovered by setting the magnetic part of the Weyl tensor to zero. Allowing it to be non-zero permits to define the dual momenta at first order. However, it is not clear how relaxing this still makes sense at non-linear order. We hope to report on this in the near future \cite{CDV}.

In the AdS linearized case, the NUT charge can be computed using the Cotton tensor \cite{Dehouck}.

\section{Supergravity}

In pure $\mathcal{N}=2$ supergravity, the charged Taub-NUT solution has a BPS bound $M^2+N^2=Q^2 +H^2$. This solution is supersymmetric as we computed in \cite{Argurio:2008zt} the globally well-defined Killing spinor \footnote{This is equivalent to the expressions given in \cite{Argurio:2008zt}  although written in a much more compact form.}:
\beqs
\epsilon= \frac{1}{2} \sqrt{\frac{r-M}{R}} e^{\frac{\beta}{2}\gamma_5}\biggr [ e^{\alpha_m \gamma_5}+i  e^{-\alpha_q \gamma_5} \gamma_0 \biggl ]  \epsilon_0(\theta, \phi)
\eeqs
where $\beta=\arctan(N/r)$, $\alpha_m=\arctan(N/M)$, $\alpha_q=\arctan(H/Q)$, $R^2=r^2+N^2$ and $\epsilon_0(\theta, \phi)$ are the flat space Killing spinors.

To recover the BPS bound from the supersymmetry algebra, we need to allow for $K_{\mu}$. Our guess consists in complexifying it by considering  the complexified Witten-Nester form. We obtain:
\beqs
\{Q,Q^*\}=\gamma^{\mu} C P_{\mu}+\gamma_5 \gamma^{\mu} C K_{\mu} -i(Q+\gamma_5 H)C
\eeqs
where we introduced a new supercharge $Q^*$. Actually, we were only able to make sense of this superalgebra when $P_{\mu}=\lambda K_{\mu}$ and in this case the new supercharge is related to the usual one by a factor $\gamma_5$, which is how the fermionic supercharge transforms when bosonic supercharges on the rhs rotate under gravitational duality. In this case, one can re-express this algebra in the usual hermitian way:
\beqs
\{Q,Q\}=\gamma^{\mu} C P'_{\mu}-i(Q'+\gamma_5 H')C
 \eeqs
This analysis shows that supersymmetry seems to be preserved under duality.

In \cite{Argurio:2008nb}, we also studied this phenomena in $\mathcal{N}=1$ supergravity where shock pp-waves are half-BPS solutions. Under gravitational duality, pp-waves whose metrics are characterized  by the knowledge of an harmonic function $F$ are sent to supersymmetric (dual) pp-waves characterized by the conjugate harmonic function of $F$. The dual of the Aichelburg-Sexl pp-wave was denoted the NUT-wave.

 For the gauged $\mathcal{N}=2$ supergravity with $\Lambda <0$,  supersymmetry of the Plebanski-Demianski solution was studied in \cite{AlonsoAlberca:2000cs}. Supersymmetric solutions with NUT charge were  also considered.  Although electromagnetic duality is broken and also apparently the gravitational duality, the BPS bound is invariant under a mixed $U(1)$ symmetry ( see also \cite{Dehouck}).


\begin{thebibliography}{9}
\bibitem{Henneaux:2004jw}
  M.~Henneaux and C.~Teitelboim,
  Phys.\ Rev.\  D {\bf 71} (2005) 024018
  [arXiv:gr-qc/0408101].

\bibitem{Deser:2005sz}
  S.~Deser and D.~Seminara,
  Phys.\ Rev.\  D {\bf 71} (2005) 081502
  [arXiv:hep-th/0503030].

\bibitem{Argurio:2009xr}
  R.~Argurio and F.~Dehouck,
  Phys.\ Rev.\  D {\bf 81}, 064010 (2010)
  [arXiv:0909.0542 [hep-th]].

\bibitem{Julia:2005ze}
  B.~Julia, J.~Levie and S.~Ray,
  JHEP {\bf 0511} (2005) 025
  [arXiv:hep-th/0507262].


\bibitem{Argurio:2008zt}
  R.~Argurio, F.~Dehouck and L.~Houart,
  Phys.\ Rev.\  D {\bf 79}, 125001 (2009)
  [arXiv:0810.4999 [hep-th]].

\bibitem{Barnich:2008ts}
  G.~Barnich and C.~Troessaert,
  JHEP {\bf 0901}, 030 (2009)
  [arXiv:0812.0552 [hep-th]].

\bibitem{Bossard:2008sw}
  G.~Bossard, H.~Nicolai and K.~S.~Stelle,
  Gen.\ Rel.\ Grav.\  {\bf 41} (2009) 1367
  [arXiv:0809.5218 [hep-th]].

\bibitem{Argurio:2008nb}
  R.~Argurio, F.~Dehouck and L.~Houart,
  JHEP {\bf 0901} (2009) 045
  [arXiv:0811.0538 [hep-th]].

\bibitem{AlonsoAlberca:2000cs}
  N.~Alonso-Alberca, P.~Meessen and T.~Ortin,
  Class.\ Quant.\ Grav.\  {\bf 17} (2000) 2783
  [arXiv:hep-th/0003071].

\bibitem{CDV}
C.~Comp\`ere, F.~Dehouck, A.~Virmani,
Work in progress

\bibitem{Dehouck}
F.~Dehouck,
Work in progress










\end{thebibliography}
\end{document}